\documentclass[aps,twocolumn,preprintnumbers,amsmath,superscriptaddress,amssymb,floats,nofootinbib]{revtex4-1}
\usepackage{soul}
\usepackage{graphicx,color}
\usepackage[colorlinks=true,citecolor=blue,linkcolor=blue]{hyperref}
\begin{document}

\title{Neutrino Oscillation Studies with Reactors}
\author{P. Vogel}\email[]{pvogel@caltech.edu}
\affiliation{Kellogg Radiation Laboratory, California Institute of Technology, Pasadena, CA 91125, USA}
\author{L. J. Wen}\email[]{wenlj@ihep.ac.cn}
\affiliation{Institute of High Energy Physics, Beijing 100049, China}
\author{C. Zhang}\email[]{chao@bnl.gov}
\affiliation{Brookhaven National Laboratory, Upton, NY 11973, USA}

\begin{abstract}
Nuclear reactors are one of the most intense, pure, controllable, cost-effective, and well-understood sources of neutrinos. Reactors have played a major
role in the study of neutrino oscillations, a phenomenon that indicates that neutrinos have mass and that neutrino flavors are quantum mechanical mixtures. Over the past several decades reactors were used in the discovery of neutrinos, were crucial in solving the solar neutrino puzzle, and allowed the determination of the smallest mixing angle $\theta_{13}$. In the near future, reactors will help to determine the neutrino mass hierarchy and to solve the puzzling issue of sterile neutrinos.
\end{abstract}

\maketitle
\thispagestyle{plain}

\section{Introduction}
\label{sec:intro}
Neutrinos, the products of  radioactive decay among other things, are somewhat enigmatic, since they can travel enormous distances through matter without interacting even once.
Understanding their properties in detail is fundamentally important. Notwithstanding that they are so very difficult to observe, great progress in this field has been achieved in recent decades.
The study of neutrinos is opening a path for the generalization of the so-called Standard Model that explains most of what we know
about elementary particles and their interactions, but in the view of most physicists is incomplete.

The Standard Model of electroweak interactions,
developed in late 1960s, incorporates neutrinos ($\nu_e$, $\nu_\mu$, $\nu_\tau$) as left-handed partners of the three families of charged leptons ($e^-$, $\mu^-$, $\tau^-$).
Since weak interactions are the only way neutrinos interact with anything, the unneeded right-handed components of the neutrino field are absent
in the Model by definition and neutrinos are assumed to be massless, with the individual lepton number (i.e. the number of leptons of a
given flavor or family) being strictly conserved.
This assignment was supported by the lack of observation of decays like
$\mu^+ \rightarrow e^+ + \gamma$ or $K_L \rightarrow e^{\pm} + \mu^{\mp}$,
despite the long search for them.

The discovery of neutrino oscillations over the past several decades
proved that these assumptions were incorrect. That discovery
represents one of the very few instances that show that
the Standard Model is indeed incomplete.
The phenomenon of neutrino oscillations means that
neutrinos have a finite mass, albeit very small, and that lepton flavor is not a conserved quantity.
Box 1 explains the basic physics of neutrino oscillations and their relation with neutrino masses, and introduces the parameters used in the oscillation formalism. Determination of all their values,
with ever increasing accuracy,
was and continues to be the main goal of neutrino experiments.
The current experimental values of the mass-squared differences $\Delta m^2_{ij}$ and the mixing angles $\theta_{ij}$ can be found
in the latest editions of the Review of Particle Physics~\cite{PDG14} and is also shown in Box 1.
Historically, the concept of neutrino oscillations was first considered by Pontecorvo~\cite{Pontecorvo57, Pontecorvo58}
and by Maki, Nakagawa and Sakata~\cite{MNS62}, hence the neutrino mixing matrix is usually called the PMNS matrix.

\begin{figure}[tb]
\begin{centering}
\includegraphics[width=\columnwidth]{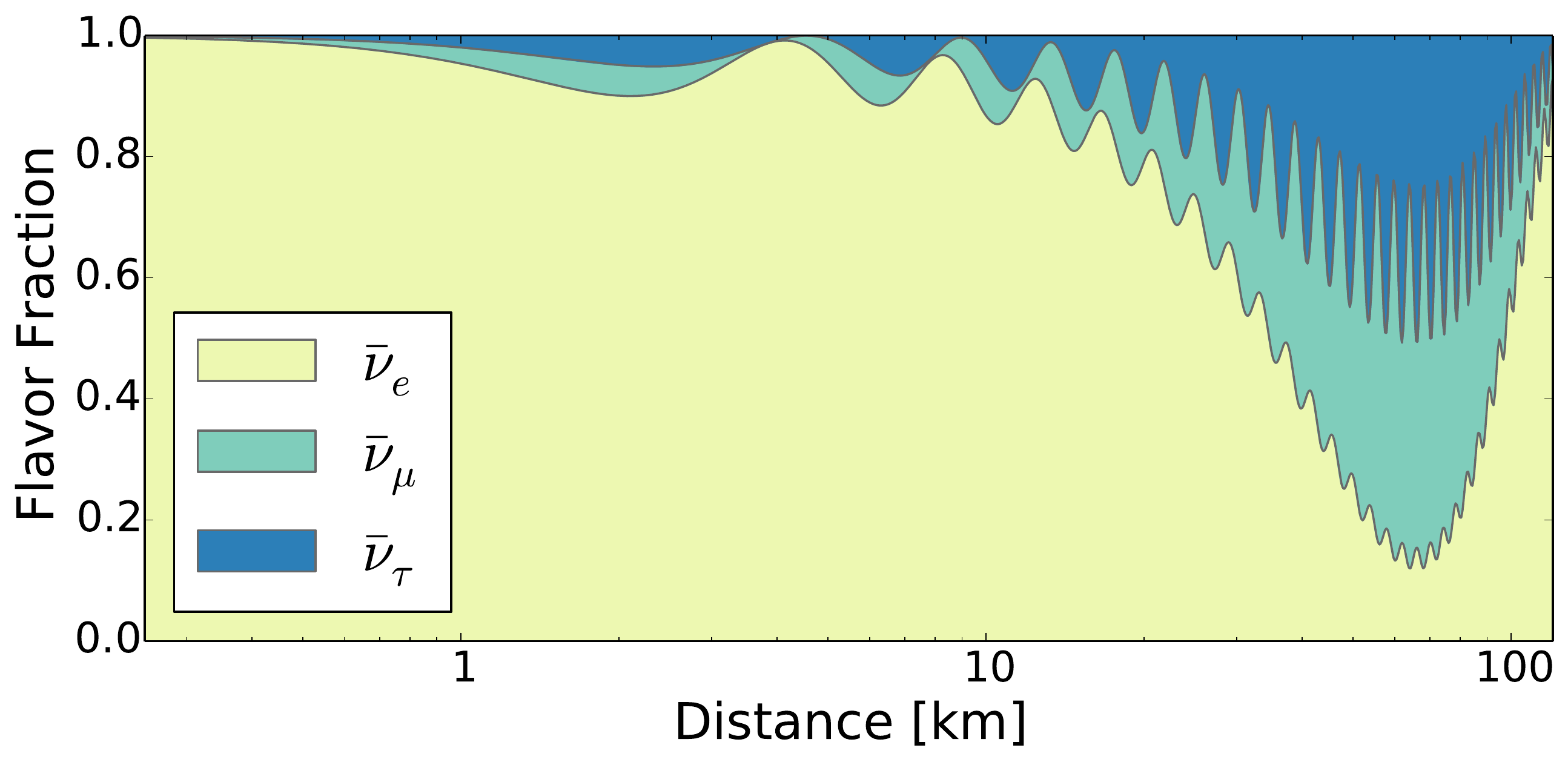}
\par\end{centering}
\caption{\label{fig:intro1} {\bf Illustration of neutrino oscillations.} The expected flavor composition of the reactor neutrino flux, for neutrinos of 4 MeV energy
used as an example, is plotted as a function of distance to the reactor cores. The fraction of neutrino flavors is calculated based on the neutrino oscillation theory introduced in Box 1. Reactor neutrino oscillation experiments are placed at different baselines to measure the oscillation features driven by different mechanisms. The experiments are usually categorized as: very short-baseline ($L\sim10$ m); short-baseline ($L\sim100$ m); kilometer-baseline ($L\sim1$ km); medium-baseline ($L\sim50$ km); and long-baseline ($L>100$ km) experiments. Details of the past and future reactor experiments, their goals and achievements, are the main topics of this Review.
}
\end{figure}

The study of reactor neutrinos played a very significant part in the discovery and detailed study of neutrino oscillations and will continue
to be essential to its further progress. Here we briefly review
the main points of this saga. Fig.~\ref{fig:intro1} illustrates how the flavor composition of the reactor neutrino flux, starting as pure $\bar\nu_e$ at production (see the next section for details), is expected to oscillate as a function of distance.
Experimental verification of this behavior, and quantitative analysis of the results, are the main topics discussed below.

The existence of neutrinos was predicted by Pauli already in 1930~\cite{Pauli30} in his famous letter attempting to explain the continuous electron energy
distribution in nuclear beta decay.
However, it was not until 1953--1959 that Reines and Cowan~\cite{Reines53,Cowan56,Reines59} were able to show that neutrinos were real particles.
Their observation used the electron antineutrinos emitted by a nuclear reactor and started a long tradition of fundamental discoveries using reactor $\bar{\nu}_e$'s.

In the early experiments detectors were placed at distances $L \le 100$ m~\cite{ILL,Gosgen,Rovno,Krasnoyarsk,SRP,Bugey4,Bugey3} (for a review see~\cite{Bemporad02}).
These pioneering short-baseline experiments, in agreement with the later established three-neutrino oscillation theory, did not observe variations with distance,
but they were important for the understanding of the reactor neutrino flux and spectrum.
The KamLAND experiment~\cite{Kamland03,Kamland05,Kamland08} in the 2000s convincingly showed that the earlier solar neutrino measurements were indeed caused by oscillations.
It demonstrated for the first time that the reactor neutrinos indeed oscillate, i.e. that the $\bar{\nu}_e$ component changes with $L/E_{\nu}$, as explained in Box 1.
It also allowed the most accurate determination of the mass-squared difference $\Delta m^2_{21}$.

In the next generation of reactor experiments including Daya Bay~\cite{Dayabay,Dayabay14}, RENO~\cite{Reno}  and Double Chooz~\cite{DChooz,DChooz14}, the longstanding puzzle of the
value of the mixing angle $\theta_{13}$ was successfully resolved; it turns out that its value $\theta_{13} \sim 8.9^\circ$ (or $\sin^2 2\theta_{13}$ = 0.093~\cite{PDG14})
is not as small as many physicists expected. That discovery opened opportunities for further experiments that should eventually
let us determine the so-far missing fundamental features of the oscillations, the neutrino mass hierarchy, and the phase $\delta_{CP}$ that characterizes the possible CP (charge and parity) violation.
The planned reactor experiments, JUNO~\cite{He-Now2014} and RENO-50~\cite{RENO-50}, promise to be an important step on this path.

Most of the oscillation results are well described by the simple three-neutrino generation hypothesis. However, there are a few anomalous indications,
the so-called reactor antineutrino anomaly~\cite{Mention} among them, that cannot be explained this way. If confirmed, they would indicate the existence of additional
fourth, or fifth, and so on neutrino families called sterile neutrinos.
These neutrinos lack weak interactions and would be observable only when mixing with the familiar active neutrinos.
The proposed very short-baseline reactor experiments at distances $\sim$10 m will test whether this fascinating possibility is realistic or not.

The discovery of neutrino oscillations is one of the most important events in the
field of particle physics at the present time. In this work we briefly review the
contribution to this achievement of the experiments involving neutrinos emitted
by nuclear reactors. First, we show how the reactor flux and its
spectrum are determined. Then, we describe the success of the
KamLAND experiment that is complementary to the exploration of solar neutrino
oscillations, and the determination of the smallest mixing angle
$\theta_{13}$ using three independent reactor experiments.

While those achievements involved considerable effort, answering the remaining
open questions is even more complicated. We begin by describing the planned large reactor experiments, JUNO and RENO-50, aiming at the difficult determination
of the so far unknown neutrino mass hierarchy or mass ordering.  We then touch on the future very short-baseline reactor experiments aiming to test the tentative and unexpected
possibility that additional light sterile neutrinos might exist.

\section{Reactor Neutrino Flux and Spectrum}
\label{sec:flux}

Nuclear reactors derive their power from fission. The fission fragments are neutron rich and undergo a cascade of $\beta$ decays.
Each fission is accompanied by approximately 6 $\beta$ decays, producing an electron and electron antineutrino each.
The decay energy, typical for the nuclear $\beta$ decay, is a few MeVs, rarely exceeding $\sim$8 MeV.
Since a typical power reactor core has thermal power of about 3 GW$_{th}$, and produces
$\sim$ 200 MeV of energy in each fission, the typical yield of $\bar{\nu}_e$ at equilibrium is $\sim 6 \times 10^{20} \bar{\nu}_e$ core$^{-1}$ s$^{-1}$.
Reactors are therefore powerful sources of low energy $\bar{\nu}_e$'s.

Neutrinos can be detected through charged current interactions when they produce charged particles, electron (mass 0.511 MeV),
muon (mass 105.7 MeV) or tau (mass 1776.8 MeV), with neutrino energy sufficient to
produce them. The reactor $\bar{\nu}_e$ energy is low, thus only reactions producing positrons are possible. Hence, to study neutrino oscillations with nuclear reactors, one must use the disappearance type of tests, i.e.~measure the flux as a function of the distance $L$ and energy $E_{\nu}$ (see Box ~1 for the detailed formalism) and
look for the deviation from the simple geometrical scaling. Traditionally such measurements were compared with the expected  $\bar{\nu}_e$ spectrum
of the reactor. Good knowledge of that spectrum, its normalization and the associated uncertainties, is essential in that case.
To reduce the dependence on the knowledge of the reactor spectrum, more recent experiments
\cite{Dayabay,Reno} use two essentially similar detectors, one nearer the reactor and another farther away.

There are two principal and complementary ways to evaluate antineutrino spectra associated with fission.
 The summation method uses known cumulative fission yields $Y_n (Z,A,t)$, and combines them
 with the experimentally known (or theoretically deduced) branching ratios $b_{n,i}(E^i_0)$ of all $\beta$-decay branches with endpoints $E^i_0$ and a
 normalized shape function of each of these many thousands of $\beta$ decays, $P_{\bar{\nu}} (E_{\bar{\nu}},E^i_0,Z)$,
 \begin{equation}
 \frac{dN}{dE_{\bar{\nu}}} = \Sigma_n Y_n (Z,A,t) \Sigma_i b_{n,i}(E^i_0) P_{\bar{\nu}} (E_{\bar{\nu}},E^i_0,Z) ~.
 \end{equation}
 There are several difficulties with this method. The branching ratios and endpoint energies are sometimes poorly known at best, in particular for
 the short-lived fragments with large $Q$ values and many branches. The individual spectrum shape functions $P_{\bar{\nu}} (E_{\bar{\nu}},E^i_0,Z)$
 require description of the Coulomb distortions including the nuclear finite size effects, weak magnetism, and radiative corrections. In addition, not all
 decays are of the allowed type. There are numerous (about 25\%) first forbidden decays involving parity change, where the individual spectrum
 shapes are much more difficult to evaluate.

 \begin{figure}[tb]
 \begin{centering}
 \includegraphics[width=\columnwidth]{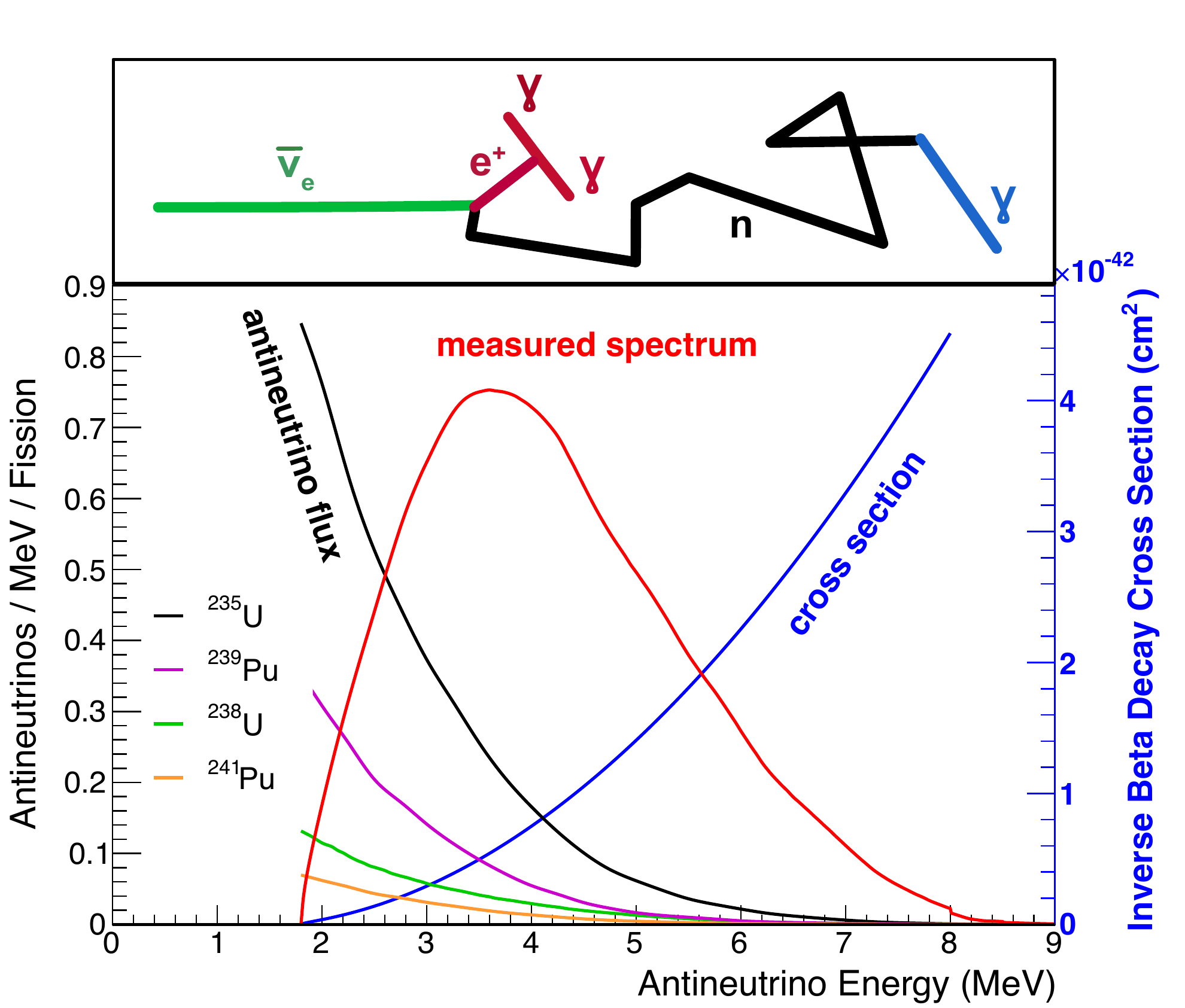}
 \par\end{centering}
 \caption{\label{fig:spectra} {\bf Detection of reactor $\bar{\nu}_e$.} In the bottom of the figure, the reactor $\bar{\nu}_e$ flux from the individual isotopes~\cite{Huber,Mueller}, weighted by their typical contribution to the total flux in a commercial reactor, is shown. The detection of $\bar{\nu}_e$ relies on the inverse beta decay reaction, whose cross section~\cite{VB99, Strumia} is shown as the blue curve. Their product is the interaction spectrum measured by the detectors, shown as the red curve.
 The steps involved in the detection are schematically drawn in the top of the figure. The $\bar{\nu}_e$ interacts with a proton, becoming a positron ($e^+$) and a neutron. The $e^+$ quickly deposits its energy and annihilates into two 511-keV $\gamma$-rays, which gives a prompt signal. The neutron scatters in the detector until being thermalized. It is then captured by a proton $\sim$ 200 $\mu$s later and releases a 2.2-MeV $\gamma$-ray (the capture time can be significantly reduced by the doping of isotopes
with very large neutron capture cross section such as gadolinium). The detection of this prompt-delayed signal pair indicates an $\bar{\nu}_e$ candidate.}
 \end{figure}

 The other method uses the experimentally determined spectrum of electrons associated with fission of the principal reactor fuels. That spectrum has been
 measured at ILL Grenoble for the thermal neutron fission of $^{235}$U, $^{239}$Pu and $^{242}$Pu~\cite{vonFeilitzsch,Schreckenbach,Hahn} and recently also for the fast neutron fission of $^{238}$U
 in Munich~\cite{Haag}. These electron spectra are then transformed into the $\bar{\nu}_e$ spectra using the obvious fact that these two leptons share the total energy
 of each $\beta$-decay branch. The transformation is on the basis of fitting first the electron spectra to a set of 30 or more virtual branches, with the equidistant
 endpoint spacing, determining from the fit their branching ratios.
 The conversion to the $\bar{\nu}_e$ spectrum is performed in each of these virtual branches. That conversion is based on the assumption
 that the electron spectrum is known precisely.
 When all virtual branches are put together one has to also take into account that different nuclear charges $Z$
 contribute with different weights to different electron
 and $\bar{\nu}_e$ energies.
 While the conversion would introduce only minimum uncertainty if all decays would be of the allowed shape, the presence
 of the first forbidden decays introduces additional uncertainty whose magnitude is difficult to determine accurately.

The summation method was used
initially in~\cite{Davis,Vogel81,Klapdor-Pu,Klapdor-U,Kopeikin} and in the more recent version in~\cite{Mueller}.
The conversion method was first used in~\cite{vonFeilitzsch,Schreckenbach,Hahn}, more details can be found in~\cite{Vogel07} and the more recent version in~\cite{Huber}.
Naturally, the thermal power of the reactor and its time-changing fuel composition must be known, as must the
energy associated with fissions of the isotopes $^{235}$U, $^{239}$Pu, $^{241}$Pu
and $^{238}$U. In addition, as already mentioned, small corrections to the spectrum shape of individual $\beta$-decay branches due to the radiative correction, weak magnetism,
nuclear size, and so on must be correctly included. Difficult to do accurately, but of a particular importance, is to take into account the spectrum shape of the numerous
first forbidden $\beta$ decays~\cite{Hayes}. The overall uncertainty in the flux was estimated in~\cite{Mueller, Huber} to be $\sim$ 2\%, but when the
first forbidden decays are included it is estimated in Ref.~\cite{Hayes} that the uncertainty increases to $\sim$ 5\%.

In essentially all reactor neutrino oscillation studies, the $\bar{\nu}_e$ are detected using the inverse neutron $\beta$-decay reaction
\begin{equation}
\bar{\nu}_e + p \rightarrow e^+ + n~, \\ ~ \sigma = 9.53 \frac{E_e p_e}{ {\rm MeV^2}} (1 + \textrm{corr.}) \times 10^{-44} {\rm cm^2} ~,
\label{eq:detection}
\end{equation}
whose cross section is accurately known~\cite{VB99, Strumia} and depends primarily on the known neutron decay half-life. (At the same time the recoil, radiative
corrections etc., must be also be taken into account.) Since the neutron is so much heavier than the available energy, its kinetic energy is quite
small (tens of keV) and thus the principal observables are the number and energy of the positrons. Most importantly, the correlated observation of the
positrons and the delayed neutron captures is a powerful tool for background suppression. Note that the reaction (\ref{eq:detection}) has
a threshold of 1.8 MeV, only $\bar{\nu}_e$ with energy larger than that can produce positrons.

In Fig.~\ref{fig:spectra} we illustrate the energy dependence of the reactor $\bar{\nu}_e$ flux, the detection reaction cross section and their product,
i.e. the measured antineutrino spectrum. The contributions of the individual isotopes to the $\bar{\nu}_e$ flux, weighted by their typical contribution
to the reactor power are also shown. The top part of the figure schematically indicates the steps involved in the $\bar{\nu}_e$ capture on proton reaction.

\onecolumngrid
\vspace{12pt}
\fboxsep=12pt

\noindent\fbox{%
   \parbox{0.96\textwidth}{%

{\Large \textcolor{blue}{Box 1: Neutrino Oscillations}}
\vspace{12pt}
\setlength{\parindent}{12pt}

Neutrinos are produced with a definite `flavor': $\nu_e, \nu_{\mu}$, or $\nu_{\tau}$. For example, in nuclear $\beta$ decay electron
antineutrinos ($\bar{\nu}_e$) are always produced together with an electron. Similarly, if a positively charged muon ($\mu^+$) is produced in the decay of the meson $\pi^+$, the muon
neutrino ($\nu_{\mu}$) is always produced as well. However, if  neutrinos have a finite
mass, the flavor composition of a neutrino beam could vary regularly as
a function of the distance and energy.  This behavior, called neutrino oscillation, is a subtle
consequence of quantum mechanics that postulates that a neutrino of a given flavor need
not be a state of a definite mass, but instead could be a coherent superposition of several states
of definite masses. Here we explain the basic ideas of this phenomenon.

Let us assume that there are only two massive neutrinos $\nu_i, i=1,2$ with different masses $m_i$.
In quantum mechanics the development of a state $\psi$ with momentum $p$  is governed
by $\psi(t) = \psi(0) e^{ipL}$. (We use the usual system of units where $\hbar = c =1$). For highly relativistic neutrinos
$p = \sqrt{E^2 - m^2} \sim E - m^2/2E$. Thus, when neutrinos propagate in vacuum over a distance $L$,
each acquires the phase $\nu_i(L) = \nu_i (0) \exp(-i m^2_i L/2E)$. (The overall phase is skipped. Additional phases are acquired when neutrinos propagate in matter, the so-called
``MSW effect''~\cite{Wolfenstein78,MS85}, which for simplicity will not be discussed here.) Assume further that the flavor neutrinos $\nu_e$ and $\nu_{\alpha}$, i.e. the neutrinos that are the partners of
charged leptons in the weak interactions, are coherent superpositions of the states $\nu_i$, i.e. $\nu_e = \cos \theta \nu_1 + \sin \theta \nu_2$, and
analogous but orthogonal combination represents the other flavor neutrino $\nu_{\alpha} = -\sin \theta \nu_1 + \cos \theta \nu_2$. These
mixtures are characterized by the parameter $\theta$, the so-called mixing angle.

Consider now a beam of neutrinos that at $L=0$ is pure $\nu_e$. Then
\begin{equation}
\nu_e (L) = \cos \theta e^{-i m_1^2 L/2E} \nu_1(0) + \sin \theta e^{-i m_2^2 L/2E} \nu_2 (0) ~.
\end{equation}
In order to observe this beam at $L$ we need to use the weak interactions.  We must therefore project the $\nu_i$ back to the flavor basis $\nu_e$ and $\nu_{\alpha}$.
Thus
\begin{equation}
\nu_e (L) = [\cos^2 \theta e^{-i m_1^2 L/2E} + \sin^2 \theta e^{-i m_2^2 L/2E}] \nu_e(0) -
\sin \theta \cos \theta [e^{-i m_1^2 L/2E} - e^{-i m_2^2 L/2E}] \nu_{\alpha}(0) ~.
\end{equation}
The probability that we detect $\nu_e$ at the distance $L$ is just the square of the corresponding $\nu_e(0)$ amplitude. The probability
of detecting $\nu_{\alpha}$ is the square of the $\nu_{\alpha}(0)$ amplitude. After some simple algebra this becomes:
\begin{equation}\label{eq:osci}
P(\nu_e \rightarrow \nu_e) = 1 - \sin^2 2 \theta \sin^2 \frac{\Delta m^2 L}{4 E} ~,~ P(\nu_e \rightarrow \nu_{\alpha}) =  \sin^2 2 \theta \sin^2 \frac{\Delta m^2 L}{4 E} ~,
\end{equation}
where $\Delta m^2 = m^2_2 - m^2_1$ is the difference of the squares of the neutrino masses.

We see that, provided $\Delta m^2 \ne 0$ and $\theta \ne 0$ or $\pi/2$, the composition of the neutrino beam oscillates as a function of $L/E_{\nu}$
with the amplitude $\sin^2 2 \theta$ and wavelength
\begin{equation}
L_{\textrm{osc}} = 4 \pi \frac{E}{\Delta m^2} ~~\equiv ~~ L_{\textrm{osc}} ({\rm m}) = \frac{2.48 E {\rm (MeV)}}{ \Delta m^2 {\rm (eV^2)}} ~.
\end{equation}

Observation of neutrino oscillations, consequently, constitutes a proof that at least some of the neutrinos have a finite mass and that the superposition is
a nontrivial one. Generalization to the realistic case of three neutrino flavors and three
states of definite mass is straightforward. The corresponding mixing is then characterized by three mixing angles $\theta_{12}, \theta_{13}, \theta_{23}$,
 one possible $CP$ violating phase $\delta_{CP}$,
and two mass square differences $\Delta m^2_{21}$ and $\Delta m^2_{32}$.
As of 2014, in the latest edition of the Review of Particle Physics~\cite{PDG14}, the best measured values using existing data are $\sin^2(2\theta_{12}) = 0.846 \pm 0.021, \, \sin^2(2\theta_{13}) = 0.093 \pm 0.008, \, \sin^2(2\theta_{23}) = 0.999^{+0.001}_{-0.018}, \, \Delta m^2_{21} = (7.53 \pm 0.18) \times 10^{-5} \textrm{eV}^2$, and $\Delta m^2_{32} = (2.44\pm 0.06) \times 10^{-3} \textrm{eV}^2$ (assuming $m_3 > m_2$). The mass ordering between $m_3$ and $m_2$, often referred to as the neutrino mass hierarchy, and the value of $\delta_{CP}$, are currently still unknown.

Neutrino oscillations have no classical analogue. They are purely quantum-mechanical phenomena, the consequence of the coherence of the
superposition of states.

   }%
}

\newpage


\twocolumngrid

\section{Exploring Solar Neutrino Oscillations on Earth}
\label{sec:kamland}

Since the late 1960s, a series of solar neutrino experiments~\cite{Homestake,GALLEX,SAGE,Kamiokande,Super-Kamiokande}, using charged current reactions, have observed a large deficit of solar $\nu_e$ flux relative to the Standard Solar Model (SSM)~\cite{Bahcall} prediction.
It appeared that more than half of the solar neutrinos were missing.
This was referred to as the ``Solar Neutrino Problem''.
In 2001, the SNO solar neutrino experiment~\cite{SNO}, for the first time, successfully measured the total flux of all three neutrino flavors ($\nu_e$, $\nu_\mu$ and $\nu_\tau$) through the neutral current channel $\nu + d \to \nu + p + n$ using heavy water as a target, that yielded results consistent with the SSM.
The SNO result is considered the ``smoking gun'' evidence of the neutrino oscillation explanation to the Solar Neutrino Problem --- the solar neutrinos, produced as electron-neutrinos from fusion and other reactions in the central region of the Sun, are transformed into other flavors when they arrive at the Earth.

\begin{figure*}[htb]
  \centering
  \includegraphics[width=0.9\textwidth]{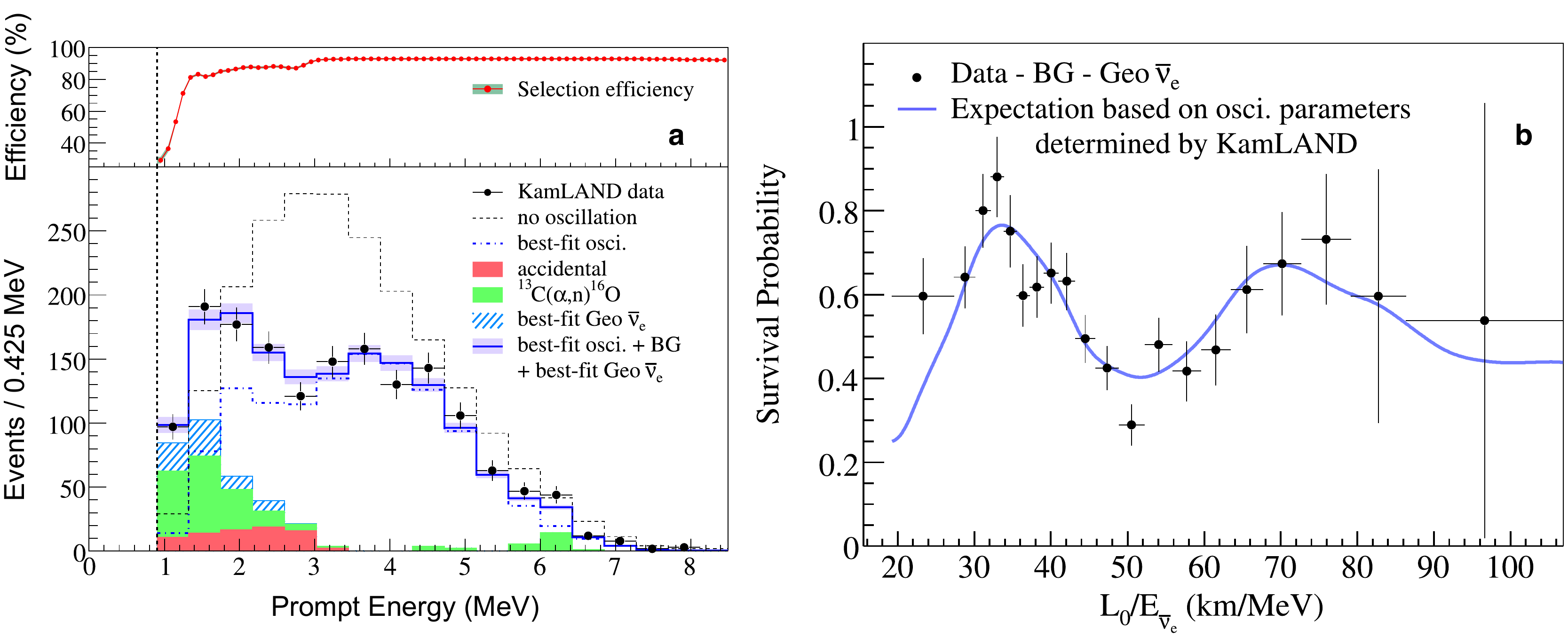}
  \caption{ \label{fig:kamland}{\bf Results from KamLAND.} (a) The data points show the measured prompt energy spectrum of $\bar\nu_e$ candidate events. The shaded histograms show the expected backgrounds. The expected reactor spectra without oscillation and with best-fit oscillation are shown as the dashed histogram. All histograms incorporate the energy-dependent selection efficiency, which is shown on the top. Only $\sim$60\% of reactor $\bar\nu_e$'s are observed relative to the no-oscillation expectation.
  (b) The data points show the ratio of the background-subtracted
  $\bar\nu_e$ spectrum to the expectation for no-oscillation as a
  function of $L_{0}/E_{\nu}$. $L_{0}$ is the effective baseline taken as a
  flux-weighted average ($L_{0}$\,=\,180\,km).
  The spectrum indicates almost two cycles of periodic feature as expected from neutrino oscillations. The oscillation survival probability using the best estimates of $\theta_{12}$ and $|\Delta{m}^2_{21}|$ is given by the blue curve.
  The curve deviates from the perfect sinusoidal $L/E$ dependence since KamLAND has multiple baselines. Figures are reproduced, with permission, from Ref.~\cite{Kamland08}.
  }
\end{figure*}

The solar neutrino experiments allowed several possible solutions in the oscillation parameter space of $\theta_{12}$ and $\Delta m^2_{21}$.
A precise measurement of these parameters and demonstration of the oscillatory feature, however, was hindered by the relatively large uncertainty of the solar $\nu_{e}$ flux predicted by the SSM, the large matter effect inside the Sun, and the extremely long distance the neutrinos travel.
A reactor neutrino experiment, measuring the same disappearance channel as the solar neutrino experiments assuming $CPT$ invariance, overcomes these difficulties.
With well-understood and controllable $\sim$MeV $\bar\nu_e$ source, a reactor experiment at $\sim$100 km baselines can explore with high precision the so-called ``Large Mixing Angle (LMA)'' parameter region suggested by the solar neutrino experiments. To do that, the KamLAND experiment~\cite{Kamland03} was built in early 2000s to explore the solar neutrino oscillations on Earth.

To shield against cosmic rays, the KamLAND detector was placed at the site of the former Kamiokande experiment~\cite{Kamiokande} under the summit of Mt.~Ikenoyama in the Japanese Alps.
The vertical overburden is 2700-m water-equivalent (m.w.e).
It is surrounded by 55 Japanese nuclear reactor cores, which then produced about 30\% of the total electricity in Japan.
The $\bar\nu_e$ flux-weighted average baseline is about 180 km, well suited for KamLAND to study the parameters suggested by the solar neutrino experiments.
The reactor operation information such as thermal power, fuel burn-up, and fuel exchange and enrichment records were provided by all Japanese reactors, which allowed KamLAND to calculate the instantaneous fission rate of each isotope accurately.

The KamLAND detector consists of 1 kton of highly purified liquid scintillator (LS), enclosed in a 13-m-diameter transparent balloon suspended by ropes in mineral oil (MO).
The MO is housed inside a 18-m-diameter stainless steel (SS) sphere, where an array of 1325 17-inch and 554 20-inch photomultiplier tubes (PMTs) is mounted.
The MO shields the inner LS region from external radiation from PMTs and SS.
Purified water (3.2 kton) is used to provide further shielding against ambient radiation and operates as an active cosmic muon veto detector.
With regular central-axis deployments of radioactive sources and dedicated off-axis deployments~\cite{KamLAND-4pi}, KamLAND achieved an excellent position resolution of $12$ cm$/\sqrt{E(\textrm{MeV})}$, energy resolution of $6.5\%/\sqrt{E(\textrm{MeV})}$, and absolute energy-scale uncertainty of 1.4\%.

Even with such powerful reactor $\bar\nu_e$ sources and a massive detector, the long baseline suppresses the expected signal at KamLAND to only about one reactor $\bar\nu_e$ event per day.
In comparison, the background from internal and external radioactivity is one million times higher.
The experiment is only possible owing to the powerful coincidence signature (the positron followed by the delayed neutron capture $\gamma$) of inverse beta decay, as illustrated in Fig.~\ref{fig:spectra}.
A time difference of less than 1 millisecond and distance less than 2 meters between the prompt and delayed signals is required in the analysis.
Only the innermost 6-m-radius scintillator region is used to reduce the accidental coincidence from external radioactivity.
Information about the event energy, position, and time were used to further reduce the accidental background to $\sim$5\% of the candidates.

The other dominant background ($\sim$10\%) at KamLAND is caused by the $^{13}$C$(\alpha,n)^{16}$O reaction where the $\alpha$-decay comes from $^{210}$Po,
a decay product of $^{222}$Rn that is naturally present in the air and many materials as traces, but is sufficient to induce a measurable contamination of the scintillator during its production.
The neutron scattering off proton or $^{16}$O$^*$ de-excitation produces a prompt signal, followed by a neutron capture delayed signal. This then mimics a true $\bar\nu_e$ event. The rest of the backgrounds include the following: the antineutrinos produced in the decay chains of $^{232}$Th and $^{238}$U in the Earth's mantle and crust, so-called geoneutrinos; the cosmogenic beta-delayed neutron emitters $^{9}$Li and $^{8}$He;  the fast neutrons from muons passing through the surrounding rock, as well as atmospheric neutrinos.

Fig.~\ref{fig:kamland} (a) shows the prompt energy spectrum of $\bar\nu_e$ candidate
events, observed with 2.9 kton$\cdot$year exposure, overlaid with the expected reactor $\bar\nu_{e}$ and background spectra. A total of 1609 events were observed, which is only about 60\% of the expected signal if there are no oscillations. The ratio of the background-subtracted $\bar\nu_e$ candidate events to no-oscillation expectation is plotted in Fig.~\ref{fig:kamland} (b) as a function of $L/E_{\nu}$. The spectrum indicates almost two cycles of the periodic feature expected from neutrino oscillations, strongly disfavoring other explanations of the $\bar\nu_e$ disappearance.

The KamLAND results~\cite{Kamland03,Kamland05,Kamland08} are highly consistent with the solar neutrino experiments, and have pinned down the solar neutrino oscillation solution to the LMA region. When combined with the results from SNO, they yield the most precise measurements of $\tan^2\theta_{12} = 0.47^{+0.06}_{-0.05}$ and $\Delta m^2_{21} = 7.59^{+0.21}_{-0.21} \times 10^{-5}$ eV$^2$.
This is a great example of the complementarity between different types of experiments. The SNO and KamLAND's first results came out within approximately 18 months of each other, with the solar experiment being more sensitive to the mixing angle $\theta_{12}$ and the reactor experiment to the mass-squared difference $\Delta m^2_{21}$. The observation of the same effect with two different sources on such different scales provides compelling evidence for neutrino oscillations.

\section{Searching for the Smallest Oscillation Angle}
\label{sec:theta13}
In contrast to the  Cabibbo-Kobayashi-Maskawa (CKM) matrix in quark mixing, where all three mixing angles are very small~\cite{PDG14}, the mixing angles in the neutrino mixing matrix appear to be large: $\theta_{23}$, measured by the atmospheric~\cite{Kajita} and long-baseline accelerator~\cite{Feldman} neutrino experiments, is consistent with $45^\circ$ which corresponds to maximal mixing; and $\theta_{12}$, measured by the solar neutrino experiments and KamLAND, is about $33^\circ$. It was therefore natural to expect that the third mixing angle, $\theta_{13}$, might be of similar magnitude.

\begin{figure*}[tb]
  \centering
  \includegraphics[width=0.95\textwidth]{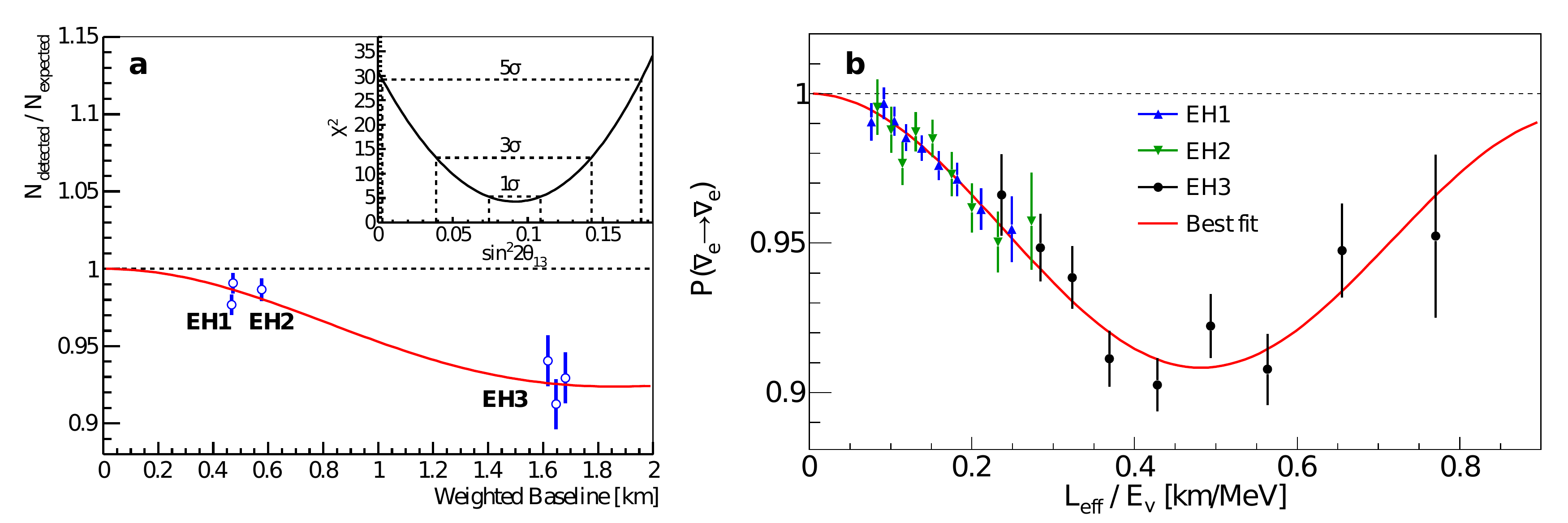}
  \caption{ \label{fig:dayabay}{\bf Results from Daya Bay.} (a) The data points show the ratio of the detected to expected no-oscillation $\bar\nu_{e}$ signals at the 6 antineutrino detectors located in three experimental halls (EHs) as a function of the weighted baseline.
  An $\sim$6\% signal deficit at the far hall relative to the near halls is observed.
  The oscillation survival probability at the best-fit value is given by the red curve. The no-oscillation hypothesis ($\theta_{13}=0$) is excluded at 5.2 standard deviations, as shown in the inset.
  (b) The data points show the ratio of the background-subtracted $\bar\nu_e$ spectrum to the expectation for no-oscillation in the three experimental halls as a function of $L_{\textrm{eff}}/E_{\nu}$. The effective baseline $L_{\textrm{eff}}$ is determined for each experimental hall (EH) equating the multi-core oscillated flux to an effective oscillated flux from a single baseline. A near-complete cycle of the expected periodic oscillation feature is observed. The oscillation survival probability using the best estimates of $\theta_{13}$ and $|\Delta{m}^2_{31}|$ is given by the red curve. Figures are reproduced, with permission, from Ref.~\cite{Dayabay} and \cite{Dayabay14}.
  }
\end{figure*}

The cleanest way to measure $\theta_{13}$ is through kilometer-baseline reactor neutrino oscillation experiments. A non-zero $\theta_{13}$ will cause a deficit of $\bar\nu_e$ flux at approximately 1--2 kilometer baseline, as indicated in Fig.~\ref{fig:intro1}. The size of the deficit is directly proportional to the value of $\sin^22\theta_{13}$.  Unlike accelerator neutrino experiments, the reactor measurements are independent of the CP phase and $\theta_{23}$, and only slightly dependent on the neutrino mass hierarchy and matter effect. A high precision measurement can therefore be achieved.

In the 1990s, two first-generation kilometer-baseline reactor experiments, CHOOZ~\cite{Chooz} and PALO VERDE~\cite{Paloverde} were constructed to measure $\theta_{13}$.
The CHOOZ detector was built at a distance of $\sim$1050 m from the two reactors of the CHOOZ power plant of \'{E}lectricit\'e de France in the Ardennes region of France. It took data from April 1997 to July 1998.
The PALO VERDE detector was built at distances of 750, 890 and 890 m from the three reactors of the Palo Verde Nuclear Generating Station in the Arizona desert of the United States. It took data between October 1998 and July 2000.
Surprisingly, neither experiment was able to observe the $\bar\nu_e$ deficit caused by $\theta_{13}$ oscillation.
As a result, only an upper limit of $\sin^22\theta_{13} < 0.10$ at 90\% C.L. was obtained~\cite{Chooz}.

The null results from CHOOZ and PALO VERDE, combined with the measured values of $\theta_{23}$ and $\theta_{12}$, motivated many phenomenological speculations of neutrino mixing patterns such as bimaximal and tribimaximal mixing~\cite{Harrison,Altarelli}.
In most of these theories, $\theta_{13}$ is either zero or very small.
A direct consequence of a vanishing $\theta_{13}$ is that the CP violation in the leptonic sector, even if large, can never be observed in the neutrino oscillation experiments.
The importance of knowing the precise value of $\theta_{13}$ provoked a series of world-wide second-generation kilometer-baseline reactor experiments in the twenty-first century, including Double Chooz~\cite{DChooz} in France, RENO~\cite{Reno} in Korea and Daya Bay~\cite{Dayabay} in China, to push the sensitivity to $\theta_{13}$ considerably below $10^\circ$.
Table~\ref{tab:theta13} summarizes some of the key parameters of the five aforementioned experiments.

\begin{table}[tb]
  \begin{tabular}{lcccc}
  \hline
  & Power & Baseline & Mass & Overburden \\
  & (GW$_{th}$) & (m) & (ton) & (m.w.e)    \\
  \hline
  CHOOZ~\cite{Chooz}            & 8.5  & 1050  & 5    & 300\\
  PALO VERDE~\cite{Paloverde}   & 11.6 & 750--890  & 12  & 32\\
  \hline
  Double Chooz~\cite{DChooz}    & 8.5  & 400  & 8  & 120\\
                                &      & 1050 & 8  & 300\\
  RENO~\cite{Reno}              & 16.8 & 290  & 16   & 120\\
                                &      & 1380 & 16   & 450\\
  Daya Bay~\cite{Dayabay}       & 17.4 & 360  & 2$\times$20   & 250\\
                                &      & 500  & 2$\times$20   & 265\\
                                &      & 1580 & 4$\times$20   & 860\\
  \hline
  \end{tabular}
  \caption{{\bf Key parameters of the reactor $\theta_{13}$ experiments.} The table summarizes the key parameters of the five past and present reactor $\theta_{13}$ experiments, including the reactor thermal power (in giga-watts), distance to reactors, target mass of the detectors, and overburden of the underground site (in meter-water-equivalent).}
\label{tab:theta13}
\end{table}

A common technology used in both the first and second generation experiments is the gadolinium-loaded liquid scintillator as the $\bar\nu_{e}$ detection target.
Gd has a high thermal neutron capture cross section. With $\sim$0.1\% gadolinium loading, the neutron capture time is reduced to $\sim$28 microseconds from $\sim$200 microseconds for the un-loaded scintillator (as used in KamLAND).
Furthermore, Gd deexcitation after the neutron capture releases an 8-MeV gamma-ray cascade, which gives a delayed signal well above natural radioactivity (in contrast,
neutron capture on a proton releases a single 2.2-MeV $\gamma$). The accidental coincidence background is therefore drastically reduced.

Addition of near detectors at baselines of a few hundred meters is the most significant improvement of the second-generation experiments over the previous ones.
As discussed above, the uncertainty in predicting the reactor antineutrino flux is relatively large (2--5\%).
This flux uncertainty, however, can be largely eliminated by the relative measurement between near and far detectors.
The Double Chooz experiment expands CHOOZ by adding a near detector at a distance of $\sim$400 m.
The installation of that near detector, however, was delayed due to civil construction.
Double Chooz started taking data in May 2011 with only a far detector, and used the Bugey4 measurement~\cite{Bugey4} to normalize the reactor flux.
The RENO experiment was built near the six reactors of the Yonggwang nuclear power plant in Korea.
The two identical detectors were located at 290 and 1380 m, respectively, from the center of the reactor array.
RENO started taking data in August 2011.
The Daya Bay experiment was built near the six reactors of the Daya Bay nuclear power plant in southern China.
Daya Bay had eight identical antineutrino detectors (ADs).
Two ADs were placed at $\sim$360 m from the two  Daya Bay reactor cores.
Two ADs were placed at $\sim$500 m from the four Ling Ao reactor cores.
Moreover, four ADs were placed at a far site $\sim$1580 m away from the 6-reactor complex.
This modular detector design allows Daya Bay to largely remove the correlated detector systematics.
Daya Bay started taking data in December 2011.

Compared with the first-generation experiments, the second-generation experiments have much larger signal statistics by utilizing higher power reactors and larger detectors. Among them, Daya Bay has the largest reactor power (17.4 GW$_{th}$) and target mass (80 tons at the far site,) as shown in Table~\ref{tab:theta13}.
The underground sites are much deeper to allow better shielding from cosmogenic background, in particular compared with the case of PALO VERDE.
Better chemical recipes of the gadolinium-loaded liquid scintillator also improve the overall detector performance and long term stability.

The second-generation reactor experiments were a huge success.
In 2012, all three experiments, Double Chooz, Daya Bay and RENO, reported
clear evidence of $\bar\nu_{e}$ disappearance at $\sim$kilometer baselines after only a few month's running~\cite{DChooz,Reno,Dayabay}.
In particular, Daya Bay excluded $\theta_{13}=0$ by 5.2 s.d.~with 55 days of data~\cite{Dayabay}. Fig.~\ref{fig:dayabay} (a) shows the result of this discovery. The ratio of the detected to expected no-oscillation $\bar\nu_{e}$ signals at the 6 detectors located in the three experimental halls is plotted as a function of weighted baseline. The signal rate at the far site shows an obvious
$\sim$6\% deficit with respect to the near sites, and fits nicely to the theoretical oscillation curve (in red). The precision of the $\theta_{13}$ measurement improved quickly with more data. With the data collected in Daya Bay through November 2013~\cite{Zhang-Neutrino14}, the best-fit value is $\sin^22\theta_{13} = 0.084 \pm 0.005$.
Although the last known, the precision in $\theta_{13}$ measurement (6\%) is now the best among all three mixing angles.

Similar to KamLAND, the ratio of the detected $\bar\nu_{e}$ events to no-oscillation expectation at Daya Bay is plotted in Fig.~\ref{fig:dayabay} (b) as a function of $L/E_{\nu}$.
The combined data from the three experimental halls show a near-complete cycle of the expected periodic oscillation feature.
The smaller amplitude and shorter wavelength of the oscillation, compared to the case of KamLAND, indicate the different oscillation component driven by $\theta_{13}$ and $\Delta{m}^2_{31}$.
The best-fit frequency of the oscillation yields $|\Delta{m}^2_{32}| = 2.39^{+0.11}_{-0.10} \times 10^{-3}$ eV$^2$ (assuming normal mass hierarchy), which is consistent and of comparable precision with the measurements of accelerator $\nu_\mu$ and $\bar\nu_\mu$ disappearance~\cite{MINOS2014,T2K2014}.
By the end of 2017, Daya Bay expects to measure both $\sin^22\theta_{13}$ and $|\Delta{m}^2_{32}|$ to precisions below 3\%~\cite{Zhang-Neutrino14}.

The discovery of $\theta_{13}$ represents another good example of the complementarity between different types of experiments. The first results from the reactor experiments~\cite{DChooz,Dayabay,Reno} and the accelerator experiments~\cite{T2K2011,MINOS2011} were released within approximately 9 months of each other, with the reactor experiments measuring $\bar\nu_e$ disappearance and the accelerator experiments measuring $\nu_e$ appearance. Seeing the same $\theta_{13}$-driven effects with different sources
of neutrinos at very different energy and baselines is a strong proof of neutrino oscillations.

The longstanding puzzle of the value of $\theta_{13}$ is now successfully resolved.
The relatively large value of $\theta_{13}$ opens the gateway for future experiments to determine the neutrino mass hierarchy and
to measure the CP-violating phase in the leptonic sector.

\section{Determination of Neutrino Mass Hierarchy}

At present only the absolute values of the neutrino mass-squared differences $\Delta m^2_{32}$ and $\Delta m^2_{31}$ are known, not their sign.
Depending on whether both $\Delta m^2_{31}$ and $\Delta m^2_{32}$ are positive or whether they are both negative the neutrino mass ordering is usually referred to as normal or inverted mass hierarchy, respectively.
The neutrino mass hierarchy (MH) is a problem of fundamental importance~\cite{MHwhitepaper} that represents an important step in the formulation of the Generalized Standard Model of particle physics.
Its determination will reduce the uncertainty in experiments aiming at the measurement of the CP-violating phase
and it will help in defining the goals of the forthcoming neutrinoless double beta decay experiments.
It will also improve our understanding of core-collapse supernovae.

The reactor $\bar\nu_e$-oscillations are modulated by terms which depend on $\Delta m^2_{31}$ and $\Delta m^2_{32}$.
At a medium baseline of $\sim$60 km, multiple small-amplitude, proportional to the $\sin^2 2 \theta_{13}$, oscillation peaks show up
on top of the long wavelength oscillation with the much larger amplitude proportional to the $\sin^2 2 \theta_{12}$, as shown in Fig.~\ref{fig:intro1}.
Depending on whether the MH is normal or inverted, the small-amplitude oscillation pattern shifts slightly.
The MH information can be extracted from this pattern by using a likelihood analysis~\cite{Li-PRD13} or the Fourier transform method~\cite{Zhan-PRD08,Zhan-PRD09}.
Additional information regarding the neutrino MH could be obtained by combining the reactor oscillation analysis with the long-baseline muon neutrino disappearance one~\cite{Minakata},
as the effective mass-squared differences measured there are different combinations of $\Delta m^2_{31}$, $\Delta m^2_{32}$ and other oscillation parameters.

\begin{figure}[tb]
  \centering
  \includegraphics[width=\columnwidth]{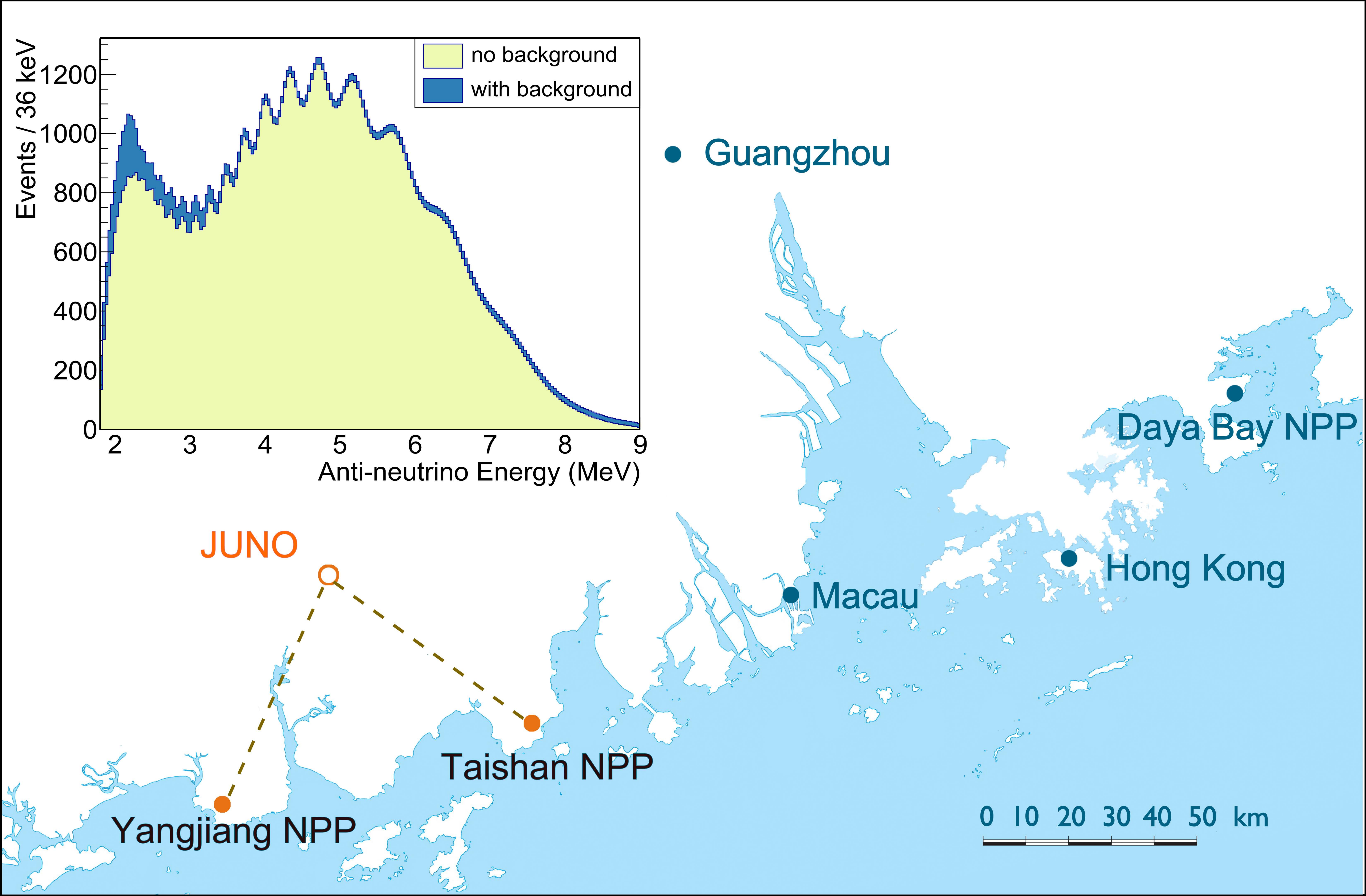}
  \caption{{\bf JUNO's layout and expected signal.} JUNO is located in southern China at an equal baseline of $\sim$53 km from the powerful reactors at Yangjiang and Taishan. The shaded histograms in the inset show the expected $\bar\nu_e$ energy spectra at JUNO with and without backgrounds after 6 years' running (calculated based on the information given in Ref.~\cite{He-Now2014}.) The multiple-oscillation structure allows the determination of neutrino mass hierarchy and precision measurement of the oscillation parameters.}
  \label{fig:juno}
\end{figure}

Two medium-baseline reactor experiments, JUNO~\cite{He-Now2014} in China and RENO-50~\cite{RENO-50} in Korea, have been proposed aiming at the MH determination, among other goals. JUNO is currently under construction.
The experiment is located in Kaiping city, Guangdong province, in southern China.
The JUNO detector will be placed underground with a total vertical overburden of 1800 m.w.e. JUNO will observe antineutrinos from the Yangjiang nuclear power plant (NPP) and the Taishan NPP at equal baselines of $\sim$53 km, as illustrated in Fig.~\ref{fig:juno}, near the maximal $\theta_{12}$-oscillation baseline. The Yangjiang NPP has six reactors cores of 2.9 GW$_{th}$ each and the Taishan NPP has planned four cores of 4.6 GW$_{th}$ each, both are under construction. The difference between the baselines to the two NPPs is controlled to less than 500 m to prevent significant degradation of the MH discrimination power~\cite{Li-PRD13}. JUNO is expected to start data taking in 2020. The proposed RENO-50~\cite{RENO-50} experiment will be located in the city of Naju, about 47 km from the Hanbit nuclear power plant with six cores of 2.8 GW$_{th}$ each. The detector will be placed at underground of Mt.~Guemseong with an overburden of 900 m.w.e. RENO-50 is expected to begin data taking in 2021.

The medium-baseline reactor experiments need to have massive detectors, approximately 20 kilotons, in order to collect sufficient $\bar\nu_e$ events in a reasonable timescale (a few years).
In the following we will primarily use JUNO as an example to illustrate the significant challenges in building such a large experiment.
The preliminary design of JUNO includes a central detector submerged in a water pool with the muon trackers installed on the top of the pool.
The water pool is equipped with PMTs and acts as an active Cherenkov detector for vetoing muons.
It also provides passive shielding against the natural radioactivities from the surrounding rock and air.
The top trackers provide complementary measurements of the cosmic muons.
The central detector consists of 20 kton of liquid scintillator (LS), contained either in a spherical acrylic tank supported by stainless steel frames, or a thin transparent balloon contained in a stainless steel vessel.
The detector looks similar to the one in SNO or KamLAND, but is twenty times larger.
In order to collect enough light, the central detector is viewed by about 18000 20-inch PMTs.
The PMTs have implosion containers to mitigate the risk of implosion chain reactions. Taking into account the mechanical clearance, the PMTs provide a near maximal surface coverage of 75\%--78\%. RENO-50 has a similar detector design with 18 kt liquid scintillator and 15000 20-inch PMTs.

An energy resolution better than $3\%/\sqrt{E(\textrm{MeV})}$ is essential for medium-baseline reactor experiments to maintain the MH discrimination ability~\cite{Li-PRD13}.
To achieve that, beside keeping a maximal photocathode coverage, additional technical improvements are necessary.
High quantum efficiency (QE $\sim$ 35\%) PMTs are necessary in order to increase the light detection efficiency. A new type of 20-inch micro-channel plate (MCP) PMTs is being developed for JUNO. The light yield and the optical transparency of the LS also need to be improved. Optimizing the concentration of scintillation fluors, purification of the raw solvent and fluors, and on-line Al$_2$O$_3$ column filtration have been found effective. LS attenuation length of more than 30~m is desired.

Calibration of the absolute energy scale is crucial.
In particular, three main effects cause non-linear energy response of a LS detector: scintillator quenching, Cherenkov radiation and possible non-linear electronics response.
If the energy non-linearity correction has large uncertainties, particular residual non-linear shapes may fake the oscillation pattern with a wrong mass hierarchy~\cite{Qian-PRD13}.
The absolute energy scale uncertainty needs to be controlled within a few tenths of a percent, which is challenging from the experience of KamLAND~\cite{KamLAND-4pi} and Daya Bay~\cite{Zhang-Neutrino14}. The requirement demands a comprehensive calibration program for a large detector such as JUNO or RENO-50.

Background control is demanding, in particular due to relatively shallow depth of the experimental sites of JUNO and RENO-50.
The sources of background are similar to those of KamLAND.
However, the cosmogenic $^9$Li and $^8$He background is significant, due to the much higher muon rate.
The $^9$Li and $^8$He isotopes are mostly produced by the muons accompanied by large electromagnetic or hadronic showers~\cite{KamLAND-spall}. In KamLAND, if a shower muon is tagged, the whole detector is vetoed for 2 s. Such a veto strategy will lead to a significant signal loss at JUNO and RENO-50.
Since the lateral distance of the isotopes from the parent muon trajectory is approximately exponential~\cite{KamLAND-spall}, a small veto region along the muon track can efficiently remove the background with minimal loss of signals. Thus, the ability to track the shower muons is essential, which demands new developments in the muon veto system and improvements on the simulations and reconstructions.

JUNO will observe about 60 reactor $\bar\nu_e$ events per day. The expected energy spectrum is shown in Fig.~\ref{fig:juno}.
The sensitivity of the mass hierarchy determination at JUNO is estimated to exceed 3$\sigma$ (for the statistical interpretations see~\cite{Qian-Stat,Blennow}) in 6 years~\cite{Li-PRD13,MBRwitepaper}.
Assuming that the effective mass-squared difference measured by the ongoing accelerator experiments can achieve 1.5--1\% precision~\cite{Agarwalla}, the MH sensitivity at JUNO can be improved to 3.7--4.4$\sigma$~\cite{Li-PRD13}.
RENO-50 has similar sensitivity reaches.
In addition to the MH determination, both JUNO and RENO-50 have great potentials in the precision measurements of the neutrino oscillation parameters. The experiments expect to measure $\Delta m^2_{21}, |\Delta m^2_{31}|$ and $\sin^2 \theta_{12}$ to precisions better than 1\%. This offers a major step toward the unitarity test of the neutrino mixing matrix~\cite{unitarity13} and is important to guide the directions of future experiments and theories.

The next-generation medium-baseline reactor experiments provide a unique opportunity to determine the neutrino mass hierarchy with the precision measurement of the reactor neutrino spectrum.
Most systematic effects are well-understood and studied, although the technical challenges are significant.
The MH sensitivity is expected to reach 3--4$\sigma$.
The reactor measurements are independent of $\theta_{23}$, the CP-violating phase, and the matter effect.
Combining with the future long-baseline accelerator~\cite{LBNE,LBNO} and atmospheric~\cite{PINGU,ORCA} neutrino oscillation programs, we will once again have complementary measurements of the neutrino mass hierarchy with different types of experiments.
Such complementarity has proved essential in the history of establishing the phenomenon of neutrino oscillations.

\section{Searching for Sterile Neutrinos}

Precision electroweak measurements of the decay width of the Z boson
determine the number of active light neutrinos. The result, $2.92 \pm 0.05$,  ~\cite{EW-2005} is obviously compatible with the three neutrino flavors.
The three-neutrino framework has been extremely successful in explaining neutrino oscillation results, since only two oscillation frequencies, corresponding to the two mass-squared differences
($\Delta m_{21}^2\sim7.6\times10^{-5}$ eV$^2$ and $\Delta m_{31}^2 \sim2.4\times10^{-3}$ eV$^2$),
were observed by the solar, atmospheric, accelerator and reactor neutrino oscillation experiments.
However, in the 1990s, the LSND experiment~\cite{LSND1995,LSND2001} reported an anomalous
event excess in the $\bar\nu_\mu\rightarrow\bar\nu_e$ appearance channel, which could be interpreted as an oscillation with the $\Delta m^2\sim1$ eV$^2$.
Such scale is clearly incompatible with the above $\Delta m_{21}^2$ and $\Delta m_{31}^2$.
Since the LSND result contradicted the three-neutrino framework, it is often referred to as the ``LSND anomaly''.

The LSND anomaly indicates the existence of additional fourth or more neutrino families with masses $m \sim$ 1 eV.
Since these additional neutrinos cannot couple to Z bosons, they must lack weak interactions and are therefore sterile.
Sterile neutrinos are observable only through their sub-dominant mixing with the familiar active neutrinos.
The light sterile neutrinos, coincidentally, are also among the leading candidates to resolve outstanding puzzles in astrophysics and
cosmology~\cite{Dodelson,Kusenko,Wyman,Battye}.
On the other hand, the light sterile neutrinos are generally not ``natural" in the theories that extend the neutrino Standard Model.
For example, the popular type-I see-saw model~\cite{Minkowski,Yanagida,GellMann,Mohapatra},
which provides an elegant explanation of the small neutrino masses and the matter-antimatter asymmetry of the universe~\cite{Fukugita}, predicts only heavy sterile neutrinos ($m >10^{10}$ eV).
If the light sterile neutrinos indeed exist, as LSND indicates, they would suggest new frontiers in both experimental and theoretical physics.

The LSND anomaly so-far remains experimentally unconfirmed, despite many efforts.
However, there are several hints supporting LSND's findings, even though none are really conclusive.
The MiniBooNE experiment, designed at a similar $L/E$ baseline as LSND using accelerator neutrinos, observed event excess
in the $\nu_{\mu}\rightarrow\nu_e$ and $\bar\nu_{\mu}\rightarrow\bar\nu_e$ appearance channels that have been interpreted as consistent with LSND~\cite{MiniBooNE2007,MiniBooNE2013}.
The GALLEX~\cite{GALLEX2010} and the SAGE~\cite{SAGE2009} solar neutrino experiments, during their calibrations using intense neutrino sources ($^{51}$Cr, $^{37}$Ar), observed a $\sim$24\% event deficit in the $\nu_e$ disappearance channel.
This deficit is often referred to as the ``Gallium anomaly''.
Recently, re-evaluations of the reactor $\bar\nu_e$ flux resulted in an increase in the predicted $\bar\nu_e$ rate~\cite{Mueller, Huber} (see earlier sections for details).
Combining the new predictions with the reactor experimental data at baselines between 10--100 m~\cite{ILL,Gosgen,Rovno,Krasnoyarsk,SRP,Bugey4,Bugey3} suggests a $\sim$4--6\% deficit between the measured and predicted reactor $\bar\nu_e$ flux, so-called
``reactor antineutrino anomaly"~\cite{Mention,Zhang13}.
These experimental anomalies can be interpreted by light sterile neutrinos~\cite{Guinti2011},
but might also be caused by an imperfect knowledge of the theoretical predictions or experimental systematics.
The preferred region ($\Delta{m}^2\sim1$ eV$^2$ and $\sin^22\theta\sim0.1$), however, is in some tension with the limits derived from other appearance~\cite{KARMEN2002,NOMAD03,OPERA13,ICARUS13} or disappearance searches~\cite{Stockdale84,Dydak84,MiniBooNE12-nubar,MiniBooNE12-nu,SuperK2000,MINOS11-NC,Bugey3,Conrad12,DayaBaySterile}.

There is, therefore, a strong motivation, despite the rather confusing present experimental status, to search for the light sterile neutrinos.
This led to a high priority world-wide program~\cite{sterileWP} with many proposed next-generation neutrino oscillation experiments.
Different technologies will be used, including short-baseline accelerator experiments~\cite{IsoDAR,OscSNS,NESSiE,LAr1-ND,nuSTORM} with various neutrino beams,
$^{51}$Cr ($^{144}$Ce-$^{144}$Pr) $\nu_e$ ($\bar\nu_e$) source experiments~\cite{Cribier2011,Dwyer2013,SOX,CeLAND} near or inside large LS detectors, as well as very short-baseline ($\sim10$ m) reactor (VSBR) $\bar\nu_e$ experiments.
In order to unambiguously resolve the LSND anomaly, the oscillation pattern in the $L/E$ space need to be observed, as in KamLAND (Fig.~\ref{fig:kamland}) and Daya Bay (Fig.~\ref{fig:dayabay}).
VSBR experiments provide unique opportunities to do so given the many advantages provided by reactors.

Multiple VSBR experiments have been proposed globally; in the U.S.~(PROSPECT~\cite{PROSPECT}, NuLat~\cite{NuLat}), France (NUCIFER~\cite{NUCIFER-2010, NUCIFER-2014}, STEREO~\cite{sterileWP}), Russia (DANSS~\cite{DANSS}, NEUTRINO-4~\cite{NEUTRINO4-2012,NEUTRINO4-2014}, POSEIDON~\cite{POSEIDON}), U.K.~(SOLID~\cite{Lasserre-Neutrino14}), and Korea (HANARO~\cite{HANARO}).
Table~\ref{tab:sterile} summarizes some of the key parameters of the proposed VSBR experiments.
The oscillation length of the $\sim$1 eV mass-scale sterile neutrinos is about 10 meters for reactor $\bar\nu_e$'s, thus all proposed experiments deploy their detectors at distances of 4-20 m from the reactor cores.
The reactor cores should preferably be compact in size to minimize the oscillations inside the cores,
so most experiments utilize compact research reactors with thermal power of tens of mega-watts. Those research reactors are typically highly enriched in $^{235}$U, in contrast to the commercial reactors in the nuclear power plants.

\begin{table}[tb]
  \begin{tabular}{lccccc}
  \hline
  & Power & Baseline & Mass & Dopant & Seg. \\
  & (MW$_{th}$) & (m) & (ton) &    & \\
  \hline
  PROSPECT~\cite{PROSPECT}  & 85  & 6--20 & 1\&10  & $^6$Li & Y \\
  NuLat~\cite{NuLat}  & 1500  & 3--8 & 1.0  & $^{10}$B, $^6$Li & Y \\
  NUCIFER~\cite{NUCIFER-2010}   & 70 & $\sim$7  & 0.7 & Gd & N \\
  STEREO~\cite{sterileWP} & 57  & $\sim$10 & 1.8  & Gd & N \\
  DANSS~\cite{DANSS} & 3000  & 9--12  & 0.9  & Gd & Y \\
  NEUTRINO-4~\cite{NEUTRINO4-2012} & 100  & 6-12  & 1.5  & Gd & N \\
  POSEIDON~\cite{POSEIDON} & 100  & 5--8  & $1.3$ & Gd & N \\
  SOLID~\cite{Lasserre-Neutrino14} & 45--80 & 6.8  & 2.9  & Gd, $^6$Li & Y \\
  HANARO~\cite{HANARO} & 30  & 6  & $\sim$1  & Gd & Y \\
  \hline
  \end{tabular}
  \caption{{\bf Key parameters of the very short-baseline reactor experiments.} The table summarizes the key parameters of the proposed very short-baseline reactor experiments, including reactor thermal power (in mega-watts), distance to reactors, target mass of the detectors, dopant material for neutron capture, and whether or not highly segmented detectors are planned.}
\label{tab:sterile}
\end{table}

Background control is a challenging task in the VSBR experiments.
The detectors are typically at shallow depth ($\sim$10 m.w.e.)\ constrained by the locations of the reactor cores.
The cosmic-ray related background is therefore high.
One advantage of using research reactors is that they can be turned on or off on demand, which helps to measure the non-reactor background.
The reactor-related backgrounds, such as fast neutrons and high energy gamma rays, are however more difficult to determine
as they appear together with the $\bar\nu_e$ signals.
Sufficient active veto and passive shielding are necessary. However, given the tight space near the reactor cores,  they have to be carefully designed.

As shown in Table~\ref{tab:sterile}, detectors are typically Gd-loaded or $^{6}$Li-loaded liquid (or solid) scintillators.
The Gd-LS technology is mature and a good pulse shape discrimination (PSD) against the neutron background has been demonstrated.
An advantage of the $^{6}$Li-loaded scintillator is that the
delayed neutron capture process $^{6}$Li$(n,\alpha)t$ produces an $\alpha$ particle and a triton, instead of a $\gamma$-ray.
This provides a good localization of the delayed signal and an additional PSD against the $\gamma$ background.
Some detectors are highly segmented into small cells in order to achieve good position resolution and further background rejection by using the multi-cell event topologies.
There are, however, more inactive layers in the segmented detectors so the edge effects have to be accurately simulated and measured.
It is also more challenging to perform calibrations and control the relative variations among cells for the segmented detectors.
For all detectors, sufficient light yield is required to precisely measure the reactor $\bar\nu_e$ spectrum and the possible distortions from neutrino oscillations.


Despite the challenges, very short-baseline reactor experiments provide a great opportunity to observe the distinctive feature of the light sterile neutrino oscillations, due to their extended range of energy (1--8 MeV) and baselines (5--20 m).
The world-wide next-generation VSBR experiments, as shown in Table~\ref{tab:sterile}, are being actively considered and pursued.
Many of them will begin taking data~\cite{Lasserre-Neutrino14} in 2015-16. Within a few years' running, they expect to cover the parameter region
suggested by the experimental anomalies with a sensitivity better than $5\sigma$ and
may tell us whether the fascinating possibility of light sterile neutrinos is true or not.

\section{Outlook}
\label{sec:final}

Over the past $\sim$60 years, nuclear reactors have proven to be one of the most powerful tools to study neutrino oscillations, the quantum-mechanical phenomenon that requires extensions to the Standard Model. Experiments at a few kilometers and at a few hundred kilometers from the reactor cores have produced some of the most convincing proofs of neutrino oscillations, by observing the oscillatory behavior of the reactor $\bar\nu_e$'s in the $L/E$ domain during their propagation. Reactor experiments measured several key parameters governing the neutrino mixing, including $\theta_{12}$, $\theta_{13}$, $\Delta{m}^2_{21}$ and $|\Delta{m}^2_{31}|$. They are essential in establishing the framework of neutrino oscillations.

Nuclear reactors will continue to help us uncover more facts about neutrinos. In the next $\sim$20 years, the upcoming next-generation reactor experiments will tell us what is the neutrino mass hierarchy and whether or not light sterile neutrinos exist. The results will have significant impact on other future programs such as neutrinoless double-beta decay experiments, long-baseline accelerator experiments, astrophysics and cosmology. Ultimately, they may hold the key to our deeper understanding of fundamental physics and the universe.

\vspace{12pt}
\section*{Acknowledgment}
We thank X.~Qian, D.~Jaffe, M.~Diwan and S.~Kettell for reading the manuscript.
The work of C.Z.~was supported in part by the Department of Energy under contracts DE-SC0012704.
The work of P.V.~was supported in part by the National Science Foundation NSF-1205977 and by the Physics Department, California Institute of Technology.
The work of L.J.W.~was supported in part by National Natural Science Foundation of China (11205183).


\end{document}